\documentclass[runningheads]{llncs}

\usepackage[T1]{fontenc}
\usepackage[utf8]{inputenc}
\usepackage{amsmath,amssymb}
\usepackage{pifont}             % \ding{55} = croix pour les propriétés non atteintes
\usepackage{graphicx}
\graphicspath{{figures/}}
\usepackage{tikz}               % pastilles (Harvey balls) du tableau de positionnement
\usepackage{float}              % [H] : place la figure système exactement où elle est déclarée
\usepackage{booktabs}
\usepackage{siunitx}            % colonnes S : alignement et format uniforme des nombres
\usepackage{multirow}
\usepackage{makecell}
\usepackage{cite}
\usepackage{url}
\usepackage[table]{xcolor}      % [table] charge colortbl : \cellcolor pour les cellules en échec (Table 4)
\usepackage{microtype}          % résorbe les micro-débordements (sûr avec llncs)
\usepackage{hyperref}
\usepackage{xurl}             % coupures dans les URL longues (note des remerciements)

\usepackage{tikz}
\usepackage{pgfplots}
\pgfplotsset{compat=1.18}
\usetikzlibrary{arrows.meta,positioning,calc,patterns,patterns.meta,shapes.misc,shapes.geometric,backgrounds}
\usepackage{amsmath}

\definecolor{figblue}{HTML}{2E5C8A}
\definecolor{figblueL}{HTML}{DCE7F2}
\definecolor{figcoral}{HTML}{C8552E}
\definecolor{figcoralL}{HTML}{F6E2D9}
\definecolor{figsand}{HTML}{8C7A4B}
\definecolor{figsandL}{HTML}{F1EBDC}
\definecolor{figgreen}{HTML}{3E7D5A}
\definecolor{figgreenL}{HTML}{E4F0E8}
\definecolor{figpurple}{HTML}{6B5B95}
\definecolor{figink}{HTML}{1E1E1E}
\definecolor{figmuted}{HTML}{5C5B57}
\definecolor{figgrid}{HTML}{E8E7E1}
\definecolor{figframe}{HTML}{9A998F}

\tikzset{
  every node/.style={font=\footnotesize, text=figink},
  sub/.style={font=\scriptsize, text=figmuted},
  blk/.style={draw, line width=0.5pt, minimum height=5.5mm, inner sep=0pt, outer sep=0pt},
  content/.style={blk, fill=figblueL, draw=figblue, text=figblue},
  anch/.style={blk, fill=figsandL, draw=figsand, text=figsand},
  noise/.style={blk, draw=figframe, pattern={Dots[distance=1.6pt, radius=0.35pt]}, pattern color=figframe},
  zero/.style={blk, draw=figframe, dashed, text=figmuted},
  span/.style={|-|, line width=0.8pt},
  flow/.style={-{Stealth[length=2mm]}, line width=0.6pt, draw=figmuted},
  proc/.style={draw=figframe, line width=0.5pt, rounded corners=1.2pt, minimum height=7.5mm, inner xsep=5pt, align=center},
  keyed/.style={proc, draw=figgreen, fill=figgreenL},
  zone/.style={draw, dashed, line width=0.6pt, rounded corners=1pt, inner sep=0pt},
}

\usepackage{pgfplots}\pgfplotsset{compat=1.18}
\usepgfplotslibrary{groupplots}
\definecolor{cTR}{HTML}{C8552E}   % Trans-RAG   coral
\definecolor{cKP}{HTML}{8C7A4B}   % Key-param   sand
\definecolor{cDC}{HTML}{2E5C8A}   % DCPE        steel blue
\definecolor{cFU}{HTML}{4E7F63}   % Fusion      muted green
\definecolor{figgrid}{HTML}{E8E7E1}\definecolor{figframe}{HTML}{9A998F}
\pgfplotsset{
  sty/.style={line width=1.0pt, mark size=1.9pt, mark options={fill=white, line width=0.8pt}},
  TR/.style={sty, cTR, mark=square, xshift=-3pt},
  KP/.style={sty, cKP, mark=triangle, dashed, mark options={fill=white, solid}, xshift=-1pt},
  DC/.style={sty, cDC, mark=o, xshift=1pt},
  FU/.style={sty, cFU, mark=diamond, xshift=3pt},
}

\newcommand{\cmark}{\ding{51}}                 % objectif atteint / revendiqué
\newcommand{\xmark}{\ding{55}}                 % propriété non fournie (mesuré)
\newsavebox{\hbfullbox}\newsavebox{\hbemptybox}\newsavebox{\hbhalfbox}
\savebox{\hbfullbox}{\tikz[baseline=-0.5ex]\fill (0,0) circle (0.42ex);}
\savebox{\hbemptybox}{\tikz[baseline=-0.5ex]\draw[line width=0.4pt] (0,0) circle (0.42ex);}
\savebox{\hbhalfbox}{\tikz[baseline=-0.5ex]{\draw[line width=0.4pt] (0,0) circle (0.42ex);%
  \fill (0,0) -- (90:0.42ex) arc (90:-90:0.42ex) -- cycle;}}
\newcommand{\hbfull}{\usebox{\hbfullbox}}
\newcommand{\hbempty}{\usebox{\hbemptybox}}
\newcommand{\hbhalf}{\usebox{\hbhalfbox}}
\definecolor{failgrey}{HTML}{ECEAE3}
\newcommand{\fail}[1]{\cellcolor{failgrey}#1}  % cellule qui échoue à la propriété (Table~\ref{tab:results})

\begin{document}

\title{Isolation Is Not Confidentiality: Security\\
Boundaries of Geometric Access Control in RAG}
\titlerunning{Isolation Is Not Confidentiality: Geometric Access Control in RAG}
% Version par defaut = LIMOS (auteurs, affiliation, remerciements) : `make` ou `pdflatex main`.
% Version anonyme de soumission : `make anon` (definit \anonymousversion -> main_anon.pdf).
\ifdefined\anonymousversion\else\def\limosversion{1}\fi
\ifdefined\limosversion
\author{Steve Nouyep \and
Maxime Puys \and
S\'ebastien Salva}
\authorrunning{S. Nouyep et al.}
\institute{Universit\'e Clermont Auvergne, CNRS, Clermont Auvergne INP, LIMOS, 63000 Clermont-Ferrand, France\\
\email{\{steve\_durand.nouyep\_noundjio, maxime.puys, sebastien.salva\}@uca.fr}}
\else
\author{Anonymous Author(s)}
\authorrunning{Anonymous}
\institute{Anonymous Institution}
\fi
\maketitle

\begin{abstract}
Retrieval-Augmented Generation (RAG) systems increasingly rely on vector
databases containing documents governed by different access-control policies.
Recent search-preserving geometric approaches enforce such access control by structuring the 
embedding space so that transformed embeddings, which we call protected representations, remain retrievable 
without exposing information across policy boundaries.

We provide a unified security analysis of this paradigm, covering policy
isolation, representation confidentiality, query forgery resistance, and
policy-membership hiding. We consider both a curious cloud service provider
observing the protected representations stored in the vector database and a curious user holding aligned pairs
of original embeddings and protected representations. Evaluating the available implementations 
of the works proposing these geometric transformation mechanisms across three 
encoders, we find that geometric transformations preserve retrieval utility and 
provide strong policy isolation, but do not provide the other security properties. 
A curious user can recover the original embeddings from a few hundred pairs and forge
policy-compatible queries at a higher cost that grows with encoder dimension
for secret per-policy transformations. Moreover, policy-dependent geometry can
leak policy membership even when unauthorized retrieval is prevented. These
results show that policy isolation, confidentiality, forgery resistance, and
membership hiding are distinct security objectives.
\end{abstract}

% --- Sections -------------------------------------------------------------
% Rôle dans le fil : poser le problème (autorisation != similarité), introduire
% le contrôle d'accès géométrique, énoncer la thèse (les propriétés de sécurité
% ne sont pas équivalentes), annoncer les contributions.
% NOTE citations : clés nommées. Groupe metadata/index/crypto = metadatarag
%   (filtering) + honeybee (index) + abe + compass (crypto). Paradigme
%   géométrique = transrag, gac. RAG/retrieval = lewis, karpukhin.
% NOTE terminologie : aligné sur "embedding" (representation confidentiality,
%   protected representations) pour cohérence avec l'abstract et le corps.
\section{Introduction}\label{sec:intro}

Retrieval-Augmented Generation (RAG) is more and more considered to ground large language
models (LLMs) with external knowledge, which is encoded as embeddings by an encoder (a model that maps text to numeric vectors) and stored in a latent 
embedding space to enable similarity search between queries and knowledge~\cite{lewis}. 
In enterprise deployments, a vector database may contain documents governed by different access-control
policies~\cite{karpukhin}. This creates a mismatch: embedding similarity search determines
which documents are relevant to a query, but not whether the user is authorized
to access them. As a result, semantically similar documents governed by
different policies may be retrieved together unless authorization is enforced
separately.

Existing approaches address this problem using metadata-based
filtering~\cite{metadatarag}, policy-specific indexes~\cite{honeybee}, or
cryptographic protection~\cite{abe,compass}. However, these approaches
either enforce authorization outside the embedding space, require separate
storage structures, or apply additional mechanisms to perform similarity search 
over encrypted embeddings. A different approach is to encode authorization directly into 
embeddings by means of geometric transformations that preserve similarity. We refer to this paradigm as \emph{search-preserving
geometric access control}. Each policy is associated with a secret transformation of the
embedding space; the transformed embedding, which we call the \emph{protected
representation}, is what the vector database stores, and authorized queries
undergo the same transformation. Recent systems implement this approach using policy-specific
transformations of stored embeddings~\cite{transrag,gac,extradim}. With policy separation
built into the embedding space, these approaches can enforce access control
directly through similarity search while preserving retrieval utility.

However, preserving retrieval does not guarantee other security properties.
Policy isolation and representation confidentiality are distinct objectives. A
transformation may prevent cross-policy retrieval during normal operation while
still being learnable from informative observations. Moreover, the same
structure used to separate policies may reveal policy membership. This raises a
broader question: what security properties does search-preserving geometric
access control actually provide?

We study this question through similarity preservation and four security properties:
policy isolation, representation confidentiality, query forgery resistance, and
policy-membership hiding. We consider two adversaries. A curious cloud service provider
has access to protected representations but no policy keys or original embeddings
for the target policy. A curious user additionally holds aligned
pairs of original embeddings and protected representations for a target policy. We use this
case to evaluate the residual protection once enough observations are available
to learn a policy-specific transformation. Our goal is therefore not to claim a
cryptographic break of a particular system, but to characterize the security
boundary of the search-preserving geometric access-control paradigm.

Our evaluation leads to three main findings. First, policy-specific geometric
transformations can preserve retrieval utility while providing strong policy
isolation. Second, isolation does not imply confidentiality or forgery
resistance: aligned observations make the evaluated transformations learnable, so
a curious user can recover the original embeddings from a few hundred pairs
and forge policy-compatible queries at a higher, dimension-dependent cost.
Third, geometric
separation can leak policy membership, even when policy isolation is
preserved.

\paragraph{Contributions.}
Our contributions are threefold:

\begin{itemize}
    \item We provide a unified security characterization of geometric access
    control for multi-policy RAG, treating retrieval utility and four security
    properties (policy isolation, representation confidentiality, query forgery
    resistance, and policy-membership hiding) as distinct objectives.

    \item We analyze the learnability of policy transformations from aligned
    pairs of original embeddings and protected representations, and distinguish representation
    confidentiality (recovering the original embedding) from query forgery
    resistance (constructing a policy-compatible query), two objectives that
    target different objects and need not hold together.

    \item We evaluate representative published mechanisms (policy-specific
    rotations, key-parametrized dimension augmentation, and
    distance-comparison-pre\-serving encryption) and their fusion, on the authors' own implementations across three
    encoders. A curious user recovers the original embeddings from a few hundred
    aligned pairs and forges policy-compatible queries at a higher,
    dimension-dependent cost, and policy-dependent
    geometric separation exposes policy membership, including for the cryptographic
    scheme.
\end{itemize}

\paragraph{Outline.}
Section~\ref{sec:related} reviews existing approaches to access control in RAG
and situates search-preserving geometric access control among them. Section~\ref{sec:threat}
defines the system and adversary models. Section~\ref{sec:analysis} presents
the security analysis of the four properties under a generic transformation
model. Section~\ref{sec:evaluation} reports the experimental evaluation of the
three mechanisms and their fusion across three encoders. Section~\ref{sec:discussion}
discusses implications and limitations, and Section~\ref{sec:conclusion}
concludes.

% Rôle dans le fil : donner le minimum de background (accès RAG, protection
% des embeddings, fuite/inversion) et positionner le travail. Chaque sous-partie
% prépare une propriété du fil (isolation / confidentialité / forgery / membership).
% NOTE citations (clés nommées) : metadata=metadatarag,rbacrag ; Trans-RAG=transrag ;
%   DCPE=dcpe,caprise ; transfos privacy=steer,shard ; inversion=song2020,vec2text ;
%   geometric AC=gac,extradim. Corrigé vs numéros bruts 14/16 et 17/18 (voir chat).
\section{Background and Related Work}
\label{sec:related}

\subsection{Access Control in RAG}

Access control in RAG systems is commonly enforced outside the embedding
space. Metadata-based filtering associates documents or chunks with
authorization attributes and restricts retrieval to those accessible to the
user~\cite{metadatarag,rbacrag,privacyrag}. Policy-specific indexes provide a
similar separation by maintaining separate retrieval structures for different
authorization domains~\cite{honeybee}, and LLM-driven or multitenant gateways
enforce authorization outside the vector database~\cite{arbiter,ogx}. These
approaches provide explicit access-control semantics but require authorization
information to remain available during retrieval.

Recent work has explored integrating access policies directly into the
embedding space~\cite{transrag,gac,extradim}. Trans-RAG applies
policy-dependent transformations to embeddings so that queries and documents
under the same policy remain compatible while different policies are
separated~\cite{transrag}; other schemes use policy-specific rotations~\cite{gac}
or key-parametrized dimension augmentation~\cite{extradim}.
This design reduces reliance on metadata filtering during vector search and
allows multiple policies to coexist within the same retrieval infrastructure.
Our work focuses on the security properties of search-preserving geometric access control rather
than on access-control enforcement alone.

\subsection{Search-Preserving Protection of Embeddings}

Related work studies transformations that protect vector representations while
preserving the similarity relationships needed for retrieval.
Distance-comparison-preserving encryption (DCPE)~\cite{dcpe} and related
constructions~\cite{caprise} aim to hide the underlying representation while
allowing a server to perform similarity-based operations. Other approaches
obscure embeddings while preserving retrieval utility, using embedding-space
alignment~\cite{steer} or cell-keyed residual splitting~\cite{shard}.

This creates an important tension. A transformation must preserve enough
structure for similarity search, but hiding that structure is precisely what
makes the original embedding difficult to recover. In particular, deterministic
transformations can become learnable from aligned pairs of original embeddings and
protected representations. Classical results already show that distance- and
property-preserving transformations are vulnerable to known-plaintext and
inference attacks that recover the transformation or the underlying
data~\cite{liu2006,turgay2008,naveed,aspekpa}. Unsupervised methods can even align
embedding spaces without paired supervision~\cite{conneau2018,grave2019}, so
aligned observations are a sufficient but not a necessary condition for learning
such a mapping. We extend this idea to the
policy-specific RAG setting and, beyond reconstruction, to query forgery and
policy-membership exposure. We therefore treat similarity preservation and representation
confidentiality as separate properties rather than assuming that one implies the
other.

\subsection{Representation Leakage and Inversion Attacks}

Learned embeddings and intermediate representations can leak information about
their inputs. Prior work has demonstrated reconstruction and inversion attacks
that recover inputs from such vectors, including attacks based on corresponding
input--representation pairs~\cite{song2020,vec2text,huang2024}. Related attacks
recover data from the access and query-leakage patterns of encrypted
nearest-neighbor search~\cite{knnleak}. These results motivate
evaluating what an adversary can learn beyond the intended retrieval operation.

Our setting differs from conventional embedding inversion because the
protected representation is generated by a policy-specific transformation. This
leads to two related but distinct attack objectives. An adversary may either
(i)~recover the original embedding from a protected representation, or (ii)~learn
the transformation well enough to construct a policy-compatible query. Query forgery
does not require recovering the original embedding and therefore cannot be
inferred from reconstruction performance alone.

\subsection{Positioning of This Work}

Prior work has studied policy-aware retrieval transformations and embedding
inversion attacks separately. RAG privacy surveys~\cite{sokragprivacy} and
provably secure designs~\cite{provsecrag} discuss related risks but do not
examine search-preserving geometric access control. We bridge these gaps by jointly
analyzing policy isolation, representation confidentiality, query forgery
resistance, and policy-membership hiding.
Table~\ref{tab:related-positioning} summarizes this positioning.

\begin{table}[H]
\centering
\caption{Positioning of this work. For each property, a circle indicates
whether it is a stated design objective of the approach
(\hbfull\ full, \hbhalf\ partial, \hbempty\ absent), while \cmark\
marks the properties evaluated in this work; $^{*}$~indicates an
open-source implementation. The circles and checkmarks do not indicate
whether a property is actually satisfied; this is determined by our
security analysis and experiments. DCPE is included as a
search-preserving cryptographic reference, not as an access-control
mechanism.}
\label{tab:related-positioning}
\footnotesize
\setlength{\tabcolsep}{2pt}
\begin{tabular}{@{}lcccc@{}}
\toprule
 & Policy & Representation & Forgery & Membership \\
Approach & isolation & confidentiality & resistance & hiding \\
\midrule
Metadata filtering~\cite{metadatarag,rbacrag} & \hbfull & \hbempty & \hbempty & \hbempty \\
Policy-specific indexes~\cite{honeybee}$^{*}$ & \hbfull & \hbempty & \hbempty & \hbempty \\
Cryptographic retrieval~\cite{abe,compass}$^{*}$ & \hbfull & \hbfull & \hbempty & \hbempty \\
DCPE~\cite{dcpe,caprise}$^{*}$ & \hbfull & \hbfull & \hbempty & \hbempty \\
Geometric access control~\cite{transrag,extradim}$^{*}$,~\cite{gac} & \hbfull & \hbhalf & \hbempty & \hbempty \\
\midrule
This work (evaluation) & \cmark & \cmark & \cmark & \cmark \\
\bottomrule
\end{tabular}
\end{table}

% Rôle dans le fil : modèle système (encodeur public, serveur de confiance,
% store non fiable) et frontière de capacité de l'adversaire (qui peut former
% des paires clair/transformé). Sépare le threat model de la méthode.
\section{System and Threat Model}
\label{sec:threat}

\subsection{System Model}
\label{sec:system-model}

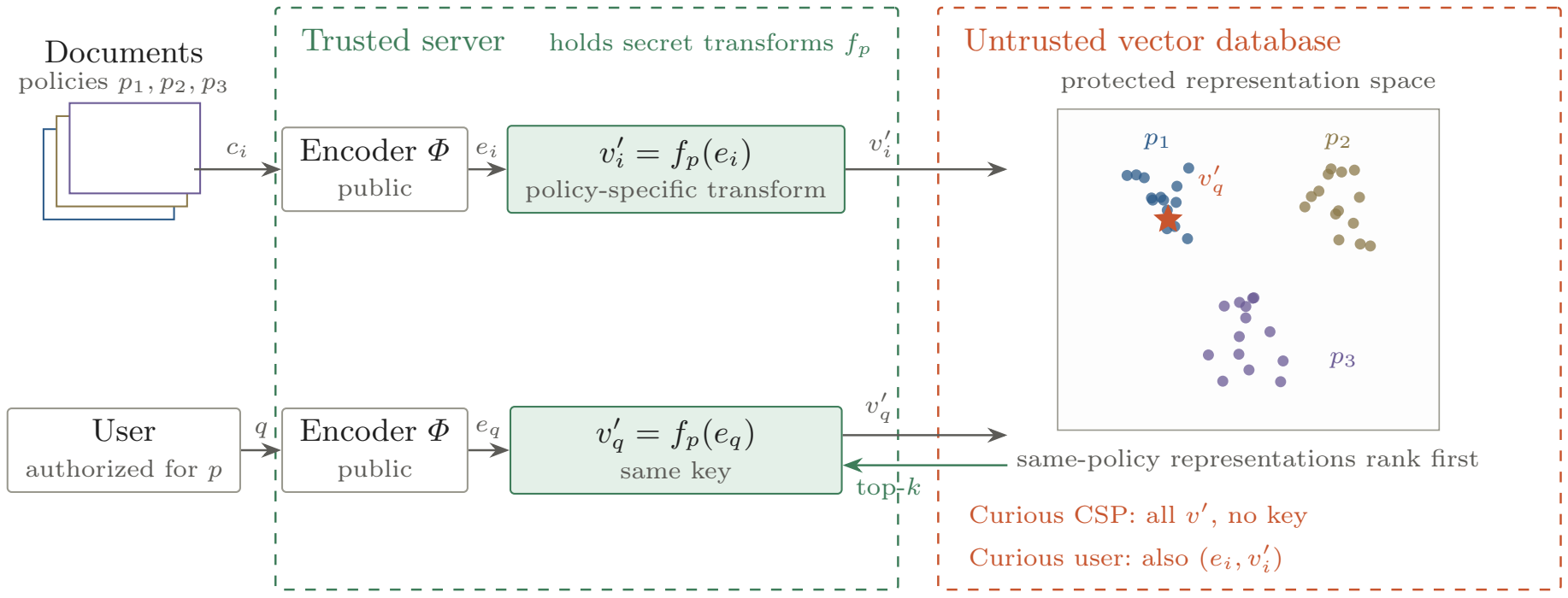
\begin{figure}[tb]
\centering
    \resizebox{\textwidth}{!}{%
        % \documentclass[tikz,border=2pt]{standalone}
% \input{figures/figstyle}
% \begin{document}
\begin{tikzpicture}[x=1mm, y=1mm]
% ---------------- trust zones
\node[zone, draw=figgreen, minimum width=62mm, minimum height=58mm, anchor=south west] (trusted) at (27, 0) {};
\node[anchor=north west, text=figgreen, font=\footnotesize] at ([xshift=1.5mm, yshift=-1mm]trusted.north west) {Trusted server};
\node[anchor=north east, text=figgreen, font=\scriptsize] at ([xshift=-1.5mm, yshift=-1.5mm]trusted.north east) {holds secret transforms $f_p$};
\node[zone, draw=figcoral, minimum width=62mm, minimum height=58mm, anchor=south west] (store) at (93, 0) {};
\node[anchor=north west, text=figcoral, font=\footnotesize] at ([xshift=1.5mm, yshift=-1mm]store.north west) {Untrusted vector database};

% ---------------- ingestion path (top)
\coordinate (y1) at (0, 42);
\foreach \i/\c in {0/figblue, 1/figsand, 2/figpurple}
  \draw[draw=\c, line width=0.5pt, fill=white] ($(4 + 1.3*\i, 37 + 1.3*\i)$) rectangle ++(13, 9);
\node[anchor=south] at (12, 51.5) {Documents};
\node[sub, anchor=south] at (12, 48.4) {policies $p_1, p_2, p_3$};
\node[proc, minimum width=17mm] (enc1) at (37, 42) {Encoder $\Phi$\\[-1pt]{\scriptsize\color{figmuted}public}};
\node[keyed, minimum width=33mm] (tr1) at (67, 42) {$v'_i = f_p(e_i)$\\[-1pt]{\scriptsize\color{figmuted}policy-specific transform}};
\draw[flow] (19, 42) -- node[above, sub] {$c_i$} (enc1.west);
\draw[flow] (enc1.east) -- node[above, sub] {$e_i$} (tr1.west);
\draw[flow] (tr1.east) -- node[above, sub, pos=0.22] {$v'_i$} (100, 42);

% ---------------- query path (bottom)
\node[proc, minimum width=17mm] (user) at (12, 14) {User\\[-1pt]{\scriptsize\color{figmuted}authorized for $p$}};
\node[proc, minimum width=17mm] (enc2) at (37, 14) {Encoder $\Phi$\\[-1pt]{\scriptsize\color{figmuted}public}};
\node[keyed, minimum width=33mm] (tr2) at (67, 14) {$v'_q = f_p(e_q)$\\[-1pt]{\scriptsize\color{figmuted}same key}};
\draw[flow] (user.east) -- node[above, sub] {$q$} (enc2.west);
\draw[flow] (enc2.east) -- node[above, sub] {$e_q$} (tr2.west);
\draw[flow] ([yshift=1.5mm]tr2.east) -- node[above, sub, pos=0.22] {$v'_q$} (100, 15.5);
\draw[flow, draw=figgreen] (100, 12.5) -- node[below, sub, text=figgreen, pos=0.72] {top-$k$} ([yshift=-1.5mm]tr2.east);

% ---------------- protected representation space (sketch)
\begin{scope}[shift={(124, 32)}]
  \draw[draw=figframe, fill=white!98!figframe, line width=0.4pt] (-19, -16) rectangle (19, 16);
  \node[sub, anchor=south] at (0, 16.5) {protected representation space};
  \pgfmathsetseed{7}
  \foreach \cx/\cy/\c/\lab/\lx/\ly in {-9/6/figblue/$p_1$/-9/12.8, 9/6/figsand/$p_2$/9/12.8, 0/-7/figpurple/$p_3$/9.5/-9} {
    \foreach \k in {1,...,14} {
      \pgfmathsetmacro{\px}{\cx + 4.2*rand}
      \pgfmathsetmacro{\py}{\cy + 4.2*rand}
      \fill[\c, opacity=0.75] (\px, \py) circle (0.55);
    }
    \node[sub, text=\c] at (\lx, \ly) {\lab};
  }
  \node[star, star points=5, star point ratio=2.2, fill=figcoral, inner sep=1.4pt] at (-8, 5) {};
  \node[text=figcoral, font=\scriptsize, anchor=south west] at (-6.2, 6.2) {$v'_q$};
  \node[sub, anchor=north] at (0, -16.8) {same-policy representations rank first};
\end{scope}

% ---------------- adversaries
\node[anchor=west, text=figcoral, font=\scriptsize] at (95, 7.5) {Curious CSP: all $v'$, no key};
\node[anchor=west, text=figcoral, font=\scriptsize] at (95, 3.2) {Curious user: also $(e_i, v'_i)$};
\end{tikzpicture}
% \end{document}
    }
\caption{System and threat model. Document and query embeddings from the public
encoder $\Phi$ are transformed by a policy-specific secret transformation $f_p$ inside the trusted
server; only the protected representations $v'$ reach the untrusted vector database, where
same-policy representations rank first. The two adversaries differ in the information
available to them: the CSP observes protected representations, whereas the
curious user additionally holds aligned pairs of original embeddings and
protected representations.}
\label{fig:system}
\end{figure}

Figure~\ref{fig:system} summarizes the system and threat model. We consider a
multi-policy RAG system in which documents governed by different
access-control policies are stored in a vector database. A public encoder
$\Phi$ maps each document chunk $c_i$ to an embedding
$e_i=\Phi(c_i)\in\mathbb{R}^{d}$. Each policy $p$ is associated with a secret
transformation $f_p$. The transformations are applied by a \emph{trusted server}
that holds the policy keys, whereas the vector database that stores the results
is untrusted. We assume that embeddings are $L_2$-normalized and that
retrieval ranks candidates by cosine similarity, i.e., the inner product of
unit-norm vectors. During ingestion, the server
applies the corresponding transformation to each embedding:

\begin{equation}
    v_i'=f_p(e_i),
\end{equation}

and stores the resulting protected representation $v_i'$ in the vector database.

At query time, an authenticated user is associated with one or more
authorized policies. For a query $q$, the server computes
$e_q=\Phi(q)$ and applies the transformation for an authorized policy before
submitting the transformed query to the vector database. The retrieved chunks
are then returned to a trusted LLM for response generation.

Policy keys never leave the server. The vector database stores only protected
representations, and similarity search is performed by a nearest-neighbor
engine (e.g., FAISS~\cite{faiss}). The provider can access these representations
but never the policy keys.

This work focuses on the security of the protected representations used for
retrieval. Document confidentiality at rest and security of the generation
component are therefore outside the scope of our analysis.

\subsection{Adversary Model}
\label{sec:adversary}

We consider honest-but-curious adversaries that follow the retrieval protocol
but may analyze protected representations available to them. All adversaries
know the public encoder $\Phi$ but do not possess the secret transformation
key of the target policy. We distinguish two levels of adversarial knowledge:

\begin{itemize}
    \item \emph{Curious Cloud Service Provider (CSP).} The provider operating the vector database
    has access to the protected representations stored in the vector database
    but no corresponding original embeddings for the target policy and no policy key.
    This models the information available to an untrusted provider.

    \item \emph{Curious User.} The adversary is a legitimately enrolled user
    that is authorized under one policy but not under the target policy. For its
    own authorized policy, it obtains pairs of original embeddings and protected
    representations through the normal system interface. For the target policy,
    we assume that it additionally obtains a bounded set of such pairs through
    collusion with the CSP; this is the most capable adversary considered in this
    work. Because the encoder $\Phi$ is public, each such pair, which we call an
    \emph{aligned pair}, provides an observation:
    \begin{equation}
        (e_i,v_i')=(\Phi(c_i),f_p(e_i)).
    \end{equation}
    These observations allow the user to estimate the policy-specific
    transformation without obtaining its secret key. This learnability from
    aligned pairs is a known weakness of distance-preserving
    transformations~\cite{liu2006,turgay2008}; we examine its implications for
    both the confidentiality of policy-specific transformations
    (Section~\ref{sec:confidentiality}) and their resistance to query forgery
    (Section~\ref{sec:forgery}).
\end{itemize}

For forgery evaluation, we additionally assume that the curious user can
submit a self-constructed protected representation of a query for similarity search under the target
policy, without possessing the target policy key or going through the trusted
server's normal query transformation. The user therefore attempts to construct
a target-policy-compatible query using the transformation estimated from the
available aligned observations. This models a compromised or colluding client
and is distinct from the passive CSP capability. This capability is not part
of normal operation; instead, it characterizes the residual protection of the
transformation once informative observations become available.

We do not consider compromise of the trusted server or disclosure of policy
keys. Active attacks on the retrieval infrastructure, such as index tampering,
denial of service, or malicious document insertion, are outside the scope of
this work. Query forgery is evaluated separately as an authorization attack
against similarity search.

\section{Security Analysis}

\label{sec:analysis}

We analyze the security properties of search-preserving geometric access
control through a component model of the underlying geometric transformations
(Section~\ref{sec:generic-model}) and under the threat model defined in
Section~\ref{sec:threat}.
We treat \emph{similarity preservation} as a functional requirement: the
transformation must preserve the similarity relationships needed for retrieval,
which we evaluate through retrieval quality relative to the unprotected
embeddings (Section~\ref{sec:evaluation}). Beyond this requirement, we study
four security properties: policy isolation, representation confidentiality,
query forgery resistance, and policy-membership hiding. A document containing
the complete derivations of this section (\texttt{docs/derivations.pdf}) is
available in the anonymized
artifact.\footnote{\url{https://anonymous.4open.science/r/geoac-rag-repro}}

\subsection{A Component Model of the Geometric Family}
\label{sec:generic-model}

We represent the considered geometric mechanisms through a common component
model. The policy-dependent components may include an orthogonal map $R_p$ (a
key-dependent coordinate permutation is a particular case, since a permutation
matrix is orthogonal), a geometric anchor $a_p\in\mathbb{R}^m$, where $m$ is the
anchor-block dimension and $m$ must be at least the number of policies for
mutually orthogonal anchors, a bounded perturbation $n$, and a possibly
nonlinear map $g$. The resulting protected
representation is:

\begin{equation}
    v' = N\!\left([\,g(R_p e+n)\,;\,M a_p\,]\right),
    \label{eq:construction}
\end{equation}

where $M$ scales the anchor and $N(\cdot)$ renormalizes the result to unit
length. Here $g$ denotes a possibly nonlinear map that some constructions add
to hinder inversion of the protected representation. It is the identity map,
denoted $g=\mathrm{id}$, for the linear constructions, whereas Trans-RAG applies
a bounded nonlinearity~\cite{transrag}. Only $R_p$ is orthogonal: once $n$, $g$, and the
renormalization are applied, the overall transformation $f_p$ is no longer an
isometry. The isolation analysis of Section~\ref{sec:isolation} therefore
takes $g=\mathrm{id}$ and is first stated without $n$; the effect of $g$ and
$n$ is evaluated empirically in Section~\ref{sec:evaluation}.

An authorized query is transformed with the same policy-specific components,
without the document-side perturbation:

\begin{equation}
    v'_q=N([\,g(R_p e_q)\,;\,M a_p\,]).
\end{equation}

We denote the noise-free linear signal-and-anchor component by
$u_p(e)=[\,R_p e\,;\,M a_p\,]$; it is the object of the isolation
analysis below.

\paragraph{Relation to existing approaches.}

The component model is not a new access-control mechanism. It is an analytical
abstraction of the published constructions (Table~\ref{tab:generic-correspondence});
all empirical attacks are performed on the published mechanisms and their fusion
(Section~\ref{sec:evaluation}).

\begin{table}[t]
\centering
\caption{Mapping of geometric components to the evaluated mechanisms
($R_p$: policy-specific orthogonal map; $M a_p$: policy anchor;
Noise: additive noise or bounded perturbation $n$; Nonlin.: nonlinear map $g$).}
\label{tab:generic-correspondence}
\small
\setlength{\tabcolsep}{6pt}
\begin{tabular}{@{}lcccc@{}}
\toprule
Mechanism & $R_p$ & $M a_p$ & Noise & Nonlin. \\
\midrule
Trans-RAG~\cite{transrag} & $\checkmark$ & -- & $\checkmark$ & $\checkmark^{\dagger}$ \\
\makecell[l]{Subspace rotation~\cite{gac}} & $\checkmark$ & -- & -- & -- \\
Key-param.~\cite{extradim} & --$^{\ddagger}$ & $\checkmark$ & -- & -- \\
DCPE~\cite{dcpe,cloakedai} & $\checkmark$ & -- & $\checkmark$ & -- \\
\bottomrule
\end{tabular}

\vspace{2pt}
\begin{flushleft}\footnotesize
$^{\dagger}$~Trans-RAG applies $g(x)=\tanh(\beta x)/\beta$ elementwise with $\beta=0.1$; the
complete mechanism is evaluated in Section~\ref{sec:evaluation}.\\
$^{\ddagger}$~Key-param.~\cite{extradim} applies a rotation shared across all policies (not
policy-specific, hence $R_p$ marked absent); only its anchor is policy-dependent.
\end{flushleft}
\end{table}

\subsection{Policy Isolation}

\label{sec:isolation}

A query authorized under policy $p$ should retrieve documents governed by $p$
rather than documents belonging to another policy. To quantify residual
cross-policy retrieval, we define two complementary rates: the \emph{query violation
rate} (QVR), the fraction of queries whose top-$k$ contains at least one
unauthorized document (\emph{how many} queries leak), and the \emph{unauthorized
chunk rate} (UCR), the fraction of retrieved chunks that are unauthorized
(\emph{how much} leaks). Both are zero under perfect isolation.

Under the cosine-similarity retrieval setting of
Section~\ref{sec:system-model}, we derive the conditions for isolation between
policies. When a policy anchor is used, as in~\cite{extradim}, policy
separation follows from its geometric contribution, and we seek the condition
on $M$ that ensures every same-policy document ranks above every cross-policy
document, regardless of content. For an authorized query $e_q$ under policy
$p$ and a document $e$ governed by policy $p$ or by another policy $p'\neq p$,
we assume unit-norm embeddings ($\|e\|=\|e_q\|=1$), an orthogonal map
($R_p^{\top}R_p=I$), unit-norm and mutually orthogonal anchors ($\|a_p\|=1$
and $\langle a_p,a_{p'}\rangle=0$ for $p\neq p'$), and $g=\mathrm{id}$. The
same-policy inner product before normalization is:

\begin{equation}
\begin{aligned}
    \langle u_p(e_q),u_p(e)\rangle
    &= \langle R_p e_q,R_p e\rangle + M^2\langle a_p,a_p\rangle \\
    &= \langle e_q,e\rangle + M^2,
\end{aligned}
\end{equation}

where we use $R_p^{\top}R_p=I$ and $\|a_p\|=1$, whereas for a cross-policy
pair it is:

\begin{equation}
\begin{aligned}
    \langle u_p(e_q),u_{p'}(e)\rangle
    &= \langle R_p e_q,R_{p'}e\rangle + M^2\langle a_p,a_{p'}\rangle \\
    &= \langle R_p e_q,R_{p'}e\rangle,
\end{aligned}
\end{equation}

since $\langle a_p,a_{p'}\rangle=0$ for $p\neq p'$.

For the noise-free idealization, in which each protected representation is
$u_p(e)$, the squared norm is:

\begin{equation}
    \|u_p(e)\|^2 = \|R_p e\|^2 + M^2\|a_p\|^2 = 1+M^2.
\end{equation}

Similarly, $\|u_p(e_q)\|^2=1+M^2$, so both cosine similarities have the same
denominator $1+M^2$:

\begin{equation}
    \operatorname{sim}_{\mathrm{same}}
    =
    \frac{\langle e_q,e\rangle+M^2}
    {1+M^2},
    \label{eq:sim-same}
\end{equation}

and,

\begin{equation}
    \operatorname{sim}_{\mathrm{cross}}
    =
    \frac{\langle R_p e_q,R_{p'}e\rangle}
    {1+M^2}.
\end{equation}

Because the embeddings are normalized,
$-1\leq\langle e_q,e\rangle\leq1$, while by the Cauchy--Schwarz inequality,

\begin{equation}
    |\langle R_p e_q,R_{p'}e\rangle|\leq1.
    \label{eq:cauchy}
\end{equation}

The common denominator therefore does not affect the relative ranking. The
same-policy numerator is at least $M^2-1$, and the cross-policy numerator is
at most $1$ by Eq.~\eqref{eq:cauchy}. Therefore, every same-policy pair ranks
above every cross-policy pair if $M^2-1>1$, which gives:

\begin{equation}
    M^2>2.
    \label{eq:isolation}
\end{equation}

The anchor is not strictly necessary for probabilistic separation. With
independently keyed, uniformly random orthogonal maps, cross-policy inner products
concentrate around zero in high dimension, with fluctuations of order
$O(1/\sqrt{d})$, where $d$ is the embedding dimension. The anchor instead provides the deterministic separation
guarantee of Eq.~\eqref{eq:isolation}. The effect of the bounded perturbation
is evaluated empirically in Section~\ref{sec:evaluation}.
This establishes the intended retrieval property, but does not imply that the
transformation or policy membership is hidden from an observer.

\subsection{Representation Confidentiality}

\label{sec:confidentiality}

The original embedding should not be readily recoverable from the protected
representation available to an adversary. We measure this property through
functional reconstruction resistance: the number of aligned pairs required to
reproduce the retrieval behavior of the original embedding
(Section~\ref{sec:evaluation}).

The component model shows that a search-preserving transformation may prevent
cross-policy retrieval while remaining learnable from sufficiently informative
observations. For the curious user, the aligned observations have the form:

\begin{equation}
    (e_i,v_i')=(e_i,f_p(e_i)),
\end{equation}

where $e_i$ is the original embedding and $v_i'$ is its protected
representation. For a linear content-dependent transformation, these pairs
constrain the policy-specific mapping. In particular, for an orthogonal
transformation $R_p$, sufficiently many linearly independent pairs allow the
mapping to be estimated by orthogonal Procrustes alignment~\cite{procrustes}.
This is analogous to known-plaintext attacks on distance-preserving
transformations~\cite{liu2006,turgay2008}, which we extend here to the RAG
access-control setting.
Thus, keeping the policy transformation secret does not by itself prevent it
from being learned. We therefore evaluate reconstruction cost under different
observation budgets in Section~\ref{sec:evaluation}.

\subsection{Query Forgery Resistance}

\label{sec:forgery}

Using the policy transformation estimated from aligned pairs
(Section~\ref{sec:threat}), the curious user should not be able to construct a
query accepted under a policy it is not authorized for. We consider
\emph{constructive forgery}: the adversary estimates the target transformation
from prior observations and applies it to a fresh query, rather than replaying
a captured legitimate query.

Forgery is evaluated separately from reconstruction because it targets a
different object: a policy-compatible query rather than an original embedding.
The forged query competes against the genuine protected representations of the target
policy, so its success depends on the accuracy of the estimated transformation,
not on the fidelity of any single reconstructed embedding.

\subsection{Policy-Membership Hiding}

\label{sec:membership}

An observer of the vector database should not be able to reliably infer which
policy a protected representation belongs to. This property is distinct from
policy isolation: a system may prevent unauthorized retrieval while still
revealing policy membership through the geometry of its representations.
Policy-dependent geometric separation can make membership observable. In
particular, policy-specific anchors give representations from the same policy a
common geometric component, while policy-specific transformations may create
additional patterns that distinguish policies.
We therefore evaluate membership inference separately from retrieval
isolation, by measuring how reliably an observer can infer the policy of a
protected representation.

% Rôle dans le fil : mesurer, sur embeddings réels, ce que chaque propriété
% vaut et confirmer la non-equivalence. Chiffres repris de la passe de
% vérification v2 (papier/v2/sections/06_evaluation.tex), sourcés artefact par
% artefact ; format condensé pour la version lite.
\section{Experimental Evaluation}

\label{sec:evaluation}

We evaluate the security properties introduced in Section~\ref{sec:analysis} on
published search-preserving geometric mechanisms and their fusion, using the
authors' own implementations. The evaluation addresses four questions:
(i) do the mechanisms preserve retrieval utility while enforcing policy
isolation? (ii) how difficult is it to recover original embeddings from
protected representations? (iii) how difficult is it to forge a query for a
policy the adversary is not authorized for? and (iv) can an observer infer
policy membership from the vector database?

\subsection{Experimental Setup}

\label{sec:setup}

We evaluate the mechanisms on the datasets and encoders summarized in
Table~\ref{tab:setup}. We run three published mechanisms with public
implementations: the query-centric transformation of Trans-RAG~\cite{transrag},
the key-parametrized dimension augmentation of~\cite{extradim}, and DCPE
through the IronCore SDK~\cite{cloakedai}. Using the authors' own
implementations ensures that the measurements characterize the published
mechanisms rather than a reinterpretation.

We additionally evaluate a \emph{fusion} that applies Trans-RAG's transformation
followed by the key-parametrized augmentation of~\cite{extradim}, to test
whether composing the two search-preserving geometric access-control mechanisms strengthens their
security properties. DCPE is evaluated separately as a distance-comparison-preserving
encryption scheme and is not part of the fusion.

Policy isolation is evaluated on a balanced MedQuAD sample with
attribute-based access-control (ABAC) metadata based on document type, project,
and region, defining nine policies. This corpus provides the multi-policy
setting needed to measure unauthorized retrieval. Retrieval utility is evaluated
separately on MS MARCO~\cite{msmarco}, a passage-ranking benchmark with
real-world search queries, to measure whether the transformations preserve
semantic retrieval quality. We additionally evaluate retrieval utility on three
BEIR datasets~\cite{beir} to test whether this behavior extends across
different retrieval corpora.

Reconstruction and forgery are evaluated on a designated target-policy subset
of MedQuAD containing 3{,}500 stored embeddings and 300 held-out queries. This
controlled single-policy setting isolates the cost of learning the target
transformation and using it for reconstruction or query forgery.

\begin{table}[t]
\centering
\caption{Evaluation corpora, tasks, and sizes. Reported sizes are the indexed
corpora and query sets. The MedQuAD ABAC sample is balanced across nine
policies (400 documents per policy); the full corpus contains 54,269 documents.}
\label{tab:setup}
\small
\begin{tabular}{@{}l l S[table-format=5.0] S[table-format=3.0] l@{}}
\toprule
Corpus & Purpose & {Docs} & {Queries} & Policies \\
\midrule
MS MARCO~\cite{msmarco} & Retrieval utility & 3047 & 306 & none \\
BEIR NFCorpus~\cite{beir} & Retrieval utility & 3633 & 323 & none \\
BEIR SciFact~\cite{beir} & Retrieval utility & 5183 & 300 & none \\
BEIR FiQA~\cite{beir} & Retrieval utility & 57638 & 648 & none \\
\midrule
MedQuAD ABAC sample~\cite{medquad} & Policy isolation & 3600 & 360 & 9 \\
MedQuAD target-policy subset & Reconstruction, forgery & 3500 & 300 & 1 \\
\bottomrule
\end{tabular}

\vspace{2pt}
{\footnotesize Encoders: MiniLM ($d{=}384$, primary), MPNet ($d{=}768$), GTE-large ($d{=}1024$).}
\end{table}

Retrieval utility is measured using Recall@10 and nDCG@10, while policy
isolation is measured using QVR and UCR (Section~\ref{sec:isolation}). Attack
success is measured functionally, by the overlap between the top-$10$ results
obtained with the estimated and true policy transformations. We denote by $n^\star_\tau$ the
smallest number of aligned pairs required to reach an overlap of $\tau$.
We use $\tau=0.8$ as the primary threshold and $\tau=0.9$ as a stricter
secondary threshold. Thus, $n^\star_\tau$ measures the empirical number of
observations required to reproduce at least $\tau$ of the reference top-$10$
results; it is not a cryptographic security guarantee.

Reconstruction and forgery use the same linear least-squares estimator. We fit
the policy transformation from the aligned pairs using the Moore--Penrose
pseudo-inverse; for the anchor-based mechanism, the constant policy anchor is
recovered from a single pair. We use the unconstrained least-squares solution
rather than an orthogonality-constrained one, so that it also accommodates
DCPE's global scaling while remaining robust to its additive perturbation.
The attack cost $n^\star_\tau$ is measured on a single witness sample using a
coarse pair grid, with 100-pair increments from 100 to 600 and larger steps
up to 3,500. Consequently,
differences such as 200 versus 300 pairs should not be interpreted as precise
one-step estimates. Isolation experiments are repeated with three random
seeds.
All experiments run on a Linux server (Ubuntu~24.04) with an Intel Xeon
Gold~5416S CPU and an NVIDIA L40 GPU. Embeddings are encoded on the GPU, while
attack-cost measurements use a single CPU thread.

\subsection{Does Geometric Transformation Preserve Retrieval and Isolation?}

\label{sec:exp-isolation}

Without policy-specific geometry, the baseline provides no policy isolation. Every
query returns at least one unauthorized document (QVR $=1.00$), while about
$8$--$9$ of the $10$ returned chunks are unauthorized (UCR $\approx0.86$).
Across the evaluated geometric configurations, both QVR
and UCR are zero, with zero variance across seeds.

On the policy-isolation corpus, geometric isolation raises authorized
within-document Recall@10 from 0.28 for the unfiltered baseline to 0.44,
matching an oracle pre-filter that removes unauthorized documents before
ranking (0.44). Policy isolation is evaluated only on the policy-labeled
MedQuAD corpus. Retrieval quality is likewise preserved across the three BEIR
collections and the three encoders: Recall@10 stays within 0.01 of the unprotected
reference in all nine dataset--encoder combinations.

Consistent with the analysis of Section~\ref{sec:isolation}, independently
keyed per-policy orthogonal maps (the rotation of Trans-RAG~\cite{transrag}
and the key-dependent coordinate permutation of DCPE~\cite{dcpe}) already
produce zero measured cross-policy
retrieval, while the anchor of the key-parametrized augmentation~\cite{extradim}
yields the same isolation (Table~\ref{tab:results}).

For retrieval utility on MS MARCO, the unprotected reference system achieves
Recall@10 $=0.986$ and nDCG@10 $=0.699$. The geometric transforms preserve
retrieval: a noise-free rotation or anchor is order-preserving, while mechanisms
that add a per-vector perturbation (Trans-RAG's blinding noise and DCPE's
perturbation) affect the ranking only marginally, keeping the top-$10$
agreement with the unprotected ranking at R@10 $\geq 0.91$ across the three encoders
(Table~\ref{tab:results}).

\paragraph{Metric sensitivity of dimension augmentation.}

The key-parametrized augmentation of~\cite{extradim} adds a large-valued
coordinate per policy, whose magnitude is calibrated for a specific similarity
metric. With the article's $\alpha=10^3$ under cosine similarity, or the code's
$\alpha=10^9$ under the Euclidean metric it targets, retrieval is preserved
(Recall@10 $\geq 0.98$ across the three encoders). Porting the code's
$\alpha=10^9$ to a cosine-based store, the common RAG configuration, collapses
retrieval (Recall@10 $=0.02$) while isolation still holds, because, with
$M=\alpha$ in Eq.~\eqref{eq:sim-same}, a difference $\Delta\cos$ between two
underlying cosine scores becomes $\Delta\cos/(1+M^2)$ and can fall below
floating-point resolution, making the resulting similarities numerically
indistinguishable. The mechanism is therefore
metric-dependent, which is a portability caveat rather than a flaw.

\subsection{Can Protected Representations Be Reconstructed?}

\label{sec:exp-reconstruction}

Table~\ref{tab:results} reports the reconstruction cost for the curious user
on the three published mechanisms and their fusion, run on the authors' own code
across three encoders of increasing dimension. The same table also reports
forgery and membership results, discussed in Sections~\ref{sec:exp-forgery}
and~\ref{sec:exp-membership}.

The reconstruction results show two main observations. First, across the three
encoders, increasing the encoder dimension from $d=384$ to $d=1024$ does not
increase the observed reconstruction cost: all evaluated transformations are
reconstructible within $200$--$400$ aligned pairs (Figure~\ref{fig:nstar}).
This is consistent with the relatively low effective dimensionality of the
embedding spaces: the participation ratio, which summarizes how many directions
carry most of the variance, is 54, 53, and 64 for the three encoders, while
90\% of the variance is captured by 159, 182, and 291 directions, respectively.
The larger encoder therefore does not raise the observed reconstruction budget;
the unprotected embeddings are likewise recovered within $300$ pairs.

Second, added noise or nonlinearity limits the \emph{fidelity} of reconstruction
but not its \emph{cost}. Trans-RAG's blinding noise and nonlinearity~\cite{transrag},
and the bounded perturbation of DCPE~\cite{dcpe}, cap the
top-$10$ overlap below one ($0.86$--$0.94$), whereas the key-parametrized
augmentation~\cite{extradim}, being a constant per-policy coordinate, leaves
reconstruction of the rotated content unimpeded (ceiling $1.00$). Yet all
reach the operational success threshold within a comparable pair budget.
Policy-specific rotations and anchors therefore do not increase the observed
reconstruction cost.

\begin{table}[t]
\centering
\caption{Published mechanisms and fusion across encoder dimensions. Reconstruction and forgery
report $n^\star_{0.8}$ and the overlap ceiling at the largest budget. R@10 is the same-policy
retrieval overlap between protected and unprotected top-$10$ results; ARI measures policy-membership clustering
by the curious CSP. Isolation is \cmark\ when QVR${=}$UCR${\approx}0$. Shaded cells mark a failed
property: isolation if QVR${}>0$; confidentiality and forgery resistance if the attack ceiling
reaches $\tau=0.8$; membership hiding if ARI${}\geq 0.5$. Pair counts $n^\star_{0.8}$ are costs,
not verdicts, and are left unshaded.}
\label{tab:results}
\footnotesize
\setlength{\tabcolsep}{4.5pt}
\begin{tabular}{@{}l S[table-format=4.0] c S[table-format=3.0] S[table-format=1.2] S[table-format=4.0] S[table-format=1.2] S[table-format=1.2] S[table-format=1.2]@{}}
\toprule
 & & & \multicolumn{2}{c}{Reconstruction} & \multicolumn{2}{c}{Forgery} & & \\
\cmidrule(lr){4-5}\cmidrule(lr){6-7}
Mechanism & {$d$} & {Iso.} & {$n^\star_{0.8}$} & {ceil.} & {$n^\star_{0.8}$} & {ceil.} & {R@10} & {ARI} \\
\midrule
No access control              & 384  & \fail{\xmark} & 300 & \fail{1.00} & \multicolumn{1}{c}{n/a} & \multicolumn{1}{c}{n/a} & 1.00 & 0.00 \\
\midrule
Trans-RAG~\cite{transrag}      & 384  & \cmark & 300 & \fail{0.86} & 500  & \fail{0.86} & 0.93 & \fail{1.00} \\
                               & 768  & \cmark & 300 & \fail{0.93} & 1000 & \fail{0.90} & 0.95 & \fail{1.00} \\
                               & 1024 & \cmark & 400 & \fail{0.87} & 1800 & \fail{0.84} & 0.91 & \fail{1.00} \\
\addlinespace
Key-param.\ aug.~\cite{extradim} & 384  & \cmark & 300 & \fail{1.00} & 300  & \fail{0.98} & 1.00 & \fail{1.00} \\
                               & 768  & \cmark & 200 & \fail{1.00} & 200  & \fail{1.00} & 1.00 & \fail{1.00} \\
                               & 1024 & \cmark & 300 & \fail{1.00} & 300  & \fail{1.00} & 1.00 & \fail{1.00} \\
\addlinespace
DCPE~\cite{dcpe,cloakedai}     & 384  & \cmark & 300 & \fail{0.90} & 500  & \fail{0.90} & 0.97 & \fail{0.97} \\
                               & 768  & \cmark & 300 & \fail{0.92} & 1000 & \fail{0.91} & 0.97 & \fail{1.00} \\
                               & 1024 & \cmark & 400 & \fail{0.91} & 1200 & \fail{0.90} & 0.96 & \fail{1.00} \\
\addlinespace
Fusion (TR${\oplus}$aug.)      & 384  & \cmark & 200 & \fail{0.86} & 500  & \fail{0.86} & 0.93 & \fail{1.00} \\
                               & 768  & \cmark & 200 & \fail{0.94} & 1000 & \fail{0.90} & 0.95 & \fail{1.00} \\
                               & 1024 & \cmark & 400 & \fail{0.88} & 1800 & \fail{0.84} & 0.91 & \fail{1.00} \\
\bottomrule
\end{tabular}
\end{table}

\subsection{Can the Same Observations Forge Policy-Compatible Queries?}

\label{sec:exp-forgery}

\begin{figure}[t]
\centering
    \resizebox{\textwidth}{!}{%
        % \documentclass[tikz,border=2pt]{standalone}
% \usepackage{pgfplots}\pgfplotsset{compat=1.18}
% \usepgfplotslibrary{groupplots}
% \definecolor{cTR}{HTML}{C8552E}   % Trans-RAG   coral
% \definecolor{cKP}{HTML}{8C7A4B}   % Key-param   sand
% \definecolor{cDC}{HTML}{2E5C8A}   % DCPE        steel blue
% \definecolor{cFU}{HTML}{4E7F63}   % Fusion      muted green
% \definecolor{figgrid}{HTML}{E8E7E1}\definecolor{figframe}{HTML}{9A998F}
% \pgfplotsset{
%   sty/.style={line width=1.0pt, mark size=1.9pt, mark options={fill=white, line width=0.8pt}},
%   TR/.style={sty, cTR, mark=square, xshift=-3pt},
%   KP/.style={sty, cKP, mark=triangle, dashed, mark options={fill=white, solid}, xshift=-1pt},
%   DC/.style={sty, cDC, mark=o, xshift=1pt},
%   FU/.style={sty, cFU, mark=diamond, xshift=3pt},
% }
% \begin{document}
\begin{tikzpicture}
\begin{groupplot}[
  group style={group size=2 by 1, horizontal sep=14mm, ylabels at=edge left},
  width=64mm, height=56mm,
  xlabel={Encoder dimension $d$}, label style={font=\footnotesize}, tick label style={font=\scriptsize},
  xtick={384,768,1024}, xmin=330, xmax=1080,
  ymin=0, ymax=1800, ytick={0,300,600,900,1200,1500,1800},
  yticklabel style={/pgf/number format/1000 sep={,}},
  grid=major, grid style={figgrid, line width=0.3pt},
  axis line style={figframe, line width=0.4pt}, axis x line*=bottom, axis y line*=left,
  title style={font=\footnotesize, at={(0,1.03)}, anchor=south west},
]
% -------- (a) reconstruction
\nextgroupplot[title={(a) Reconstruction $v'\!\to e$}, ylabel={Attack cost $n^\star_{0.8}$ (aligned pairs)},
  legend style={draw=none, fill=none, font=\scriptsize, at={(0.03,0.97)}, anchor=north west, row sep=-1.5pt},
  legend cell align=left]
\addplot[TR] coordinates {(384,300) (768,300) (1024,400)}; \addlegendentry{Trans-RAG}
\addplot[KP] coordinates {(384,300) (768,200) (1024,300)}; \addlegendentry{Key-param.\ aug.}
\addplot[DC] coordinates {(384,300) (768,300) (1024,400)}; \addlegendentry{DCPE}
\addplot[FU] coordinates {(384,200) (768,200) (1024,400)}; \addlegendentry{Fusion (TR$\oplus$aug.)}
% -------- (b) forgery
\nextgroupplot[title={(b) Query forgery $e\!\to v'_q$}]
\addplot[TR] coordinates {(384,500) (768,1000) (1024,1800)};
\addplot[KP] coordinates {(384,300) (768,200)  (1024,300)};
\addplot[DC] coordinates {(384,500) (768,1000) (1024,1200)};
\addplot[FU] coordinates {(384,500) (768,1000) (1024,1800)};
\end{groupplot}
\end{tikzpicture}
% \end{document}
    }
\caption{Attack cost $n^\star_{0.8}$ versus encoder dimension, with one line per
mechanism (values from Table~\ref{tab:results}; the same symbol denotes a
mechanism in both panels). (a)~Reconstruction stays within 200--400 aligned
pairs as $d$ grows from 384 to 1024. (b)~Query forgery grows with $d$ for mechanisms whose
per-policy transformation is secret (Trans-RAG, DCPE, and their fusion),
reaching 1,200--1,800 pairs, while the key-parametrized augmentation, whose
rotation is shared across policies, stays at 200--300. Markers are slightly
horizontally offset where values coincide.}
\label{fig:nstar}
\end{figure}
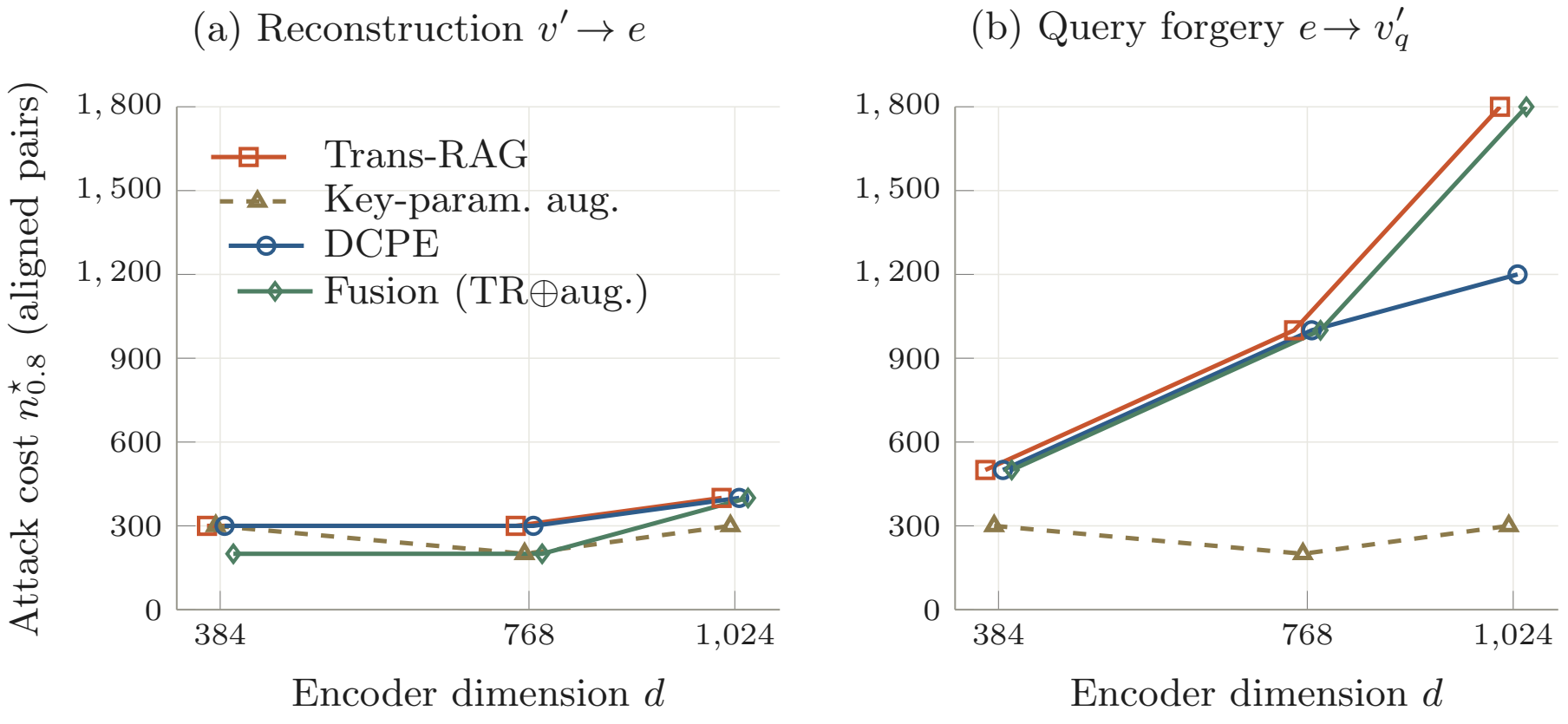

The aligned pairs used for reconstruction also let the curious user estimate the
policy transformation and forge a query under a policy it is not authorized for.
Table~\ref{tab:results} reports forgery cost alongside reconstruction. Forgery
reaches $0.84$--$1.00$ of the legitimate top-$10$ overlap across the three
published mechanisms and their fusion, so none resists the attack. Its cost,
however, differs from reconstruction in how it scales with encoder dimension.
For mechanisms whose target-policy transformation is secret (Trans-RAG, DCPE,
and the Trans-RAG$\oplus$augmentation fusion), the attacker must estimate the
full target transformation from the aligned pairs. Forgery cost therefore grows
with encoder dimension, from $500$ to $1{,}800$ pairs as $d$ increases from
$384$ to $1024$ (Figure~\ref{fig:nstar}).

The key-parametrized augmentation~\cite{extradim} behaves differently: it applies
a fixed rotation shared across all policies and separates policies only by a
constant per-policy anchor. Because the same rotation appears in every policy's
pairs, an adversary enrolled in any policy can estimate it from its own
authorized pairs, while only the low-dimensional anchor remains target-specific.
The reported $n^\star$ therefore counts the pairs needed from the adversary's own
policy; for the target policy, a single pair suffices to recover the target
anchor (Section~\ref{sec:setup}). Forgery thus does not scale with $d$ and stays at $200$--$300$ pairs, the
same cost as reconstruction.
For the mechanisms with a secret target-policy transformation, forgery costs
more than reconstruction ($500$ versus $300$ pairs at $d=384$). This higher cost is consistent
with the stricter requirement imposed by forgery: the estimated transformation
must produce a query that retrieves the target policy's genuine protected representations.

\subsection{Does Geometric Isolation Leak Policy Membership?}

\label{sec:exp-membership}

The curious CSP clusters the stored protected representations using $k$-means, without any aligned
pair or policy key. The Adjusted Rand Index (ARI) measures agreement between the
inferred clusters and the true policy partition. Across the three encoders,
Trans-RAG, the key-parametrized augmentation~\cite{extradim}, and their fusion
reach ARI $=1.00$, while DCPE reaches ARI $=0.97$--$1.00$, compared with
ARI $\approx0.001$ for the unprotected baseline (Table~\ref{tab:results}). Thus,
policy membership is highly recoverable even without aligned pairs or policy
keys, including for DCPE, a cryptographic distance-comparison-preserving scheme.

This exposure is consistent with embedding anisotropy. Learned embeddings
concentrate around a common mean, so applying a distinct per-policy key produces
distinct per-policy geometry, whether through a rotated or permuted mean for
the orthogonal-map mechanisms or through an added coordinate for the anchor.
These differences remain separable by unsupervised clustering. The same
geometric separation that supports policy isolation can therefore expose policy
membership even when unauthorized retrieval is prevented.

\subsection{Summary of Findings}

\label{sec:exp-summary}

Across the evaluated mechanisms, policy isolation does not imply reconstruction
resistance, forgery resistance, or policy-membership hiding. All geometric
mechanisms achieve the measured isolation, yet policy membership remains
recoverable from the protected representations. Increasing encoder dimension does
not increase the observed reconstruction cost, while forgery cost increases
with dimension for mechanisms with a target-policy-specific transformation.
Adding noise or nonlinearity limits reconstruction fidelity but not the observed
pair budget, and the evaluated fusion does not increase the attack costs.

The unprotected baseline provides a complementary reference: it fails isolation
(QVR${=}1.00$/UCR${=}0.86$) but does not expose policy membership
(ARI${=}0.00$). The geometric mechanisms reverse these two properties:
they provide measured isolation while exposing policy membership.

\section{Discussion}

\label{sec:discussion}

The results clarify the security boundary of search-preserving geometric access control. Geometric transformations can
enforce policy separation directly in the embedding space, but policy isolation does not imply confidentiality,
forgery resistance, or policy-membership hiding. Once the curious user holds aligned pairs, the evaluated mechanisms remain
learnable for both reconstruction and query forgery, while their policy-dependent geometry can expose membership
information.

\subsection{Limitations and Security Boundary}

The reconstruction and forgery results assume that the curious user can obtain aligned pairs of original embeddings and protected representations for the target policy. This provides more information than is normally available to the CSP and is therefore best
understood as a test of the residual protection once informative observations are available, rather than as an
assumption that such observations are always available in practice.
The evaluation covers the published mechanisms considered in this work and their fusion, not the full space of
search-preserving constructions. The measured attack costs therefore characterize these mechanisms under the evaluated
threat model and do not constitute cryptographic security guarantees for the broader paradigm.
Our formal analysis of isolation takes $g=\mathrm{id}$; the nonlinearity of Trans-RAG is retained in the empirical
evaluation but is not modeled analytically.
Overall, the results support treating policy isolation as a distinct security property rather than as a proxy for
broader confidentiality or anti-forgery guarantees. The required protection therefore depends on the security objective
and the adversarial capabilities being considered.

\subsection{Towards Mitigations}

The results also suggest several directions for mitigation. First, query
authorization should not rely solely on the geometric transformation: a trusted
server could authenticate transformed queries before they reach the vector
database, preventing a user who has learned the transformation from directly
submitting forged queries. Second, limiting the number of observations
available under a single policy transformation, for example through periodic
re-keying or stricter separation of protected representations, may reduce the
effectiveness of attacks based on aligned pairs. These measures address different parts
of the problem and should be evaluated explicitly in future work. When
representation confidentiality is a primary requirement, cryptographic
retrieval mechanisms may also be preferable to relying on geometric separation
alone.

\section{Conclusion}

\label{sec:conclusion}

Search-preserving geometric access control can provide effective policy
isolation by embedding policy separation into the embedding space. Our
analysis and experiments show that this isolation does not imply representation
confidentiality, query forgery resistance, or policy-membership hiding. With
aligned pairs of original embeddings and protected representations, a curious user can recover
those embeddings from a few hundred observations and forge policy-compatible queries
at a higher, dimension-dependent cost for secret per-policy transformations.
Composing published mechanisms does not
substantially increase these costs.
These results show that search-preserving geometric access control should be
evaluated beyond
retrieval effectiveness and policy isolation. Future work will evaluate
mitigation strategies and extend the analysis to broader classes of
search-preserving constructions.

\paragraph{Ethical statement.}

This work evaluates search-preserving geometric access-control mechanisms using
controlled experiments, publicly available data, and implementations released as
part of scientific publications. Our evaluation is limited to research
implementations intended as proofs of concept; to our knowledge, none of the
implementations evaluated in this work is known to be deployed as part of a
commercial product. The attacks target experimental representations and retrieval
interfaces in these research implementations. No real-world system was
compromised and no sensitive user data were accessed. In the event of acceptance,
we plan to contact the authors of the evaluated implementations following a
responsible-disclosure process before the final publication of the work. The
reproducibility artifact is described below.

\paragraph{Open Science.}

All code, dataset-preparation scripts, and configurations required to reproduce
the reported results, including the corpus statistics of
Table~\ref{tab:setup} and the reconstruction, forgery, and membership costs of
Table~\ref{tab:results}, are available as an anonymized artifact:
\url{https://anonymous.4open.science/r/geoac-rag-repro}. The repository provides
scripts for reproducing the main experiments, generating the reported tables and
figures, and evaluating the considered geometric mechanisms and adversarial
procedures. The artifact also contains a document with the complete derivations
of Section~\ref{sec:analysis}. Long-term archival storage is provided through
Zenodo.

% Bloc actif uniquement dans la version LIMOS (par defaut) ; absent de `make anon`.
\ifdefined\limosversion
\smallskip\noindent\textbf{Acknowledgement.}
This work was supported by the Industrial Chair on Reliable and Confident Use of
LLMs\footnote{\url{https://uca-fondation.fr/chaires/chaire-sur-lutilisation-fiable-et-de-confiance-des-grands-modeles-de-langage-llms/}}
and by the MIAI Cluster, France 2030 (ANR-23-IACL-0006).
\fi

% --- Références -------------------------------------------------------------
% ============================================================================
%  Références — bibliographie manuelle (thebibliography), sans BibTeX.
%  Toutes les entrées sont vérifiées (papier, DOI, ACL Anthology, arXiv ou
%  programme de conférence). Entrées reprises de la v1 (papier/article/
%  references.tex) après revérification le 02/09/2026.
% ============================================================================

\end{document}